\documentclass[fleqn,twoside]{article}
\usepackage{espcrc2}
\usepackage{graphicx}
\usepackage{mathrsfs}
\usepackage{amssymb,amsmath}

\def\dmatm{\Delta m_{\rm a}^2}
\def\evolt{{\rm eV}}

\title{Radiative leptogenesis in minimal seesaw models\thanks{Based on the work done
in collaboration with R. Gonz\'alez Felipe and B. Nobre
\cite{GonzalezFelipe:2003fi}.}}

\author{F. R. Joaquim\address{Dipartimento di Fisica ``G.
Galilei'', Universit\`a di Padova and
\\
\hspace{0.15cm}INFN, Sezione di Padova,\\\hspace{0.01cm} Via Marzolo, 8 - I-35131 Padua - Italy }%
 \thanks{This work was supported by {\em Funda\c c\~ao
 para a Ci\^encia e a Tecnologia} under the grant
 SFRH/BPD/14473/2003 and by INFN. } }

\begin{document}

\begin{abstract}
In the framework of seesaw models with only two heavy Majorana
neutrinos, nonzero leptonic asymmetries can be radiatively generated
when exact heavy neutrino mass degeneracy ($M_1=M_2=M$) is imposed
at a scale $\Lambda_D > M$. For a specific case, we show that an
acceptable value for the baryon asymmetry of the Universe can be
obtained considering thermal leptogenesis.
\end{abstract}

\maketitle %
Recently, a lot of attention has been payed to the study of a
possible connection between thermal leptogenesis and neutrino
masses~\cite{Giudice:2003jh}, mixing and leptonic $CP$
violation~\cite{Branco:2001pq}. For instance, it is now known that
an acceptable value for the baryon asymmetry of the Universe (BAU)
requires the low-energy Majorana neutrinos to be lighter than
$0.12-0.15$~eV. Moreover, successful thermal leptogenesis implies
$M_1>10^8-10^9$~GeV~\cite{Giudice:2003jh,Davidson:2002qv}, where
$M_1$ denotes the mass of the lightest heavy Majorana neutrino.
Although interesting, these bounds are model-dependent in the sense
that they are only valid if the heavy Majorana neutrino masses are
hierarchical.

In supersymmetric theories, the above constraint on $M_1$ may be in
conflict with the upper bound on the reheating temperature of the
Universe, which can be as low as
$10^6$~GeV~\cite{Kawasaki:2004yh}. %Such upper bounds arise from the
%requirement that the predictions of Big Bang nucleosynthesis are not
%spoiled by an overproduction of gravitinos. For a gravitino mass in
%the range $0.1-3$~TeV the upper bound on $T_{\rm RH}$ varies between
%$3\times 10^6$~GeV and $7\times 10^5$~GeV~\cite{Kawasaki:2004yh}.
%
%The aforementioned problems can be easily
This tension between the bounds on $M_1$ and $T_{\rm RH}$ can be
relaxed if one considers quasi-degenerate heavy Majorana neutrinos.
In this case, acceptable values for the BAU can be obtained with
heavy masses as low as 1~TeV~\cite{Pilaftsis:2003gt}.

In Ref.~\cite{GonzalezFelipe:2003fi} we have studied the case where
the small heavy Majorana neutrino mass-splitting, needed to enhance
the $CP$-asymmetries, is generated radiatively. For simplicity, we
have restricted ourselves to a seesaw model with only two heavy
right-handed neutrinos~\cite{Frampton:2002qc}. We denote the Dirac
neutrino and charged-lepton Yukawa coupling matrices respectively by
$Y$ and $Y_\ell\,$, while $M_R$ will stand for the $2\times2$
symmetric mass matrix of the heavy right-handed neutrinos.

We start by considering that at a scale $\Lambda_D$ the heavy
neutrinos are degenerate, i.e. $M_1=M_2 \equiv M$, with $M <
\Lambda_D$. In this limit, $CP$ is not necessarily conserved.
Indeed, the non-vanishing of the weak-basis invariant $
\mathcal{J}_1=M^{-6}{\rm Tr}\left[ Y Y^{T}
Y_{\ell}^{}Y_{\ell}^{\dagger} Y^{*}
Y^{\dagger},Y_{\ell}^{*}Y_{\ell}^{T}\right] ^{3},\label{highCPinv} $
which is not proportional to $M_{2}^2-M_{1}^2$, would signal a
violation of $CP$. On the other hand, a non-zero leptonic asymmetry
can be generated if and only if the $CP$-odd invariant $
\mathcal{J}_{2}={\rm Im\,Tr}\,[H M_R^{\dagger} M_R M_R^{\dagger}H^T
M_R ]$ does not vanish~\cite{Pilaftsis:2003gt}. Since
$\mathcal{J}_{2}$ can be expressed in the form
\begin{equation}
\mathcal{J}_{2} =M_1 M_2 (M_2^2-M_1^2)\, {\rm
Im\,}[H_{12}^2]\,\,,\,\,H=Y^\dag Y\,,
\end{equation}
$\mathcal{J}_{2}\neq 0$ requires not only $M_1 \neq M_2$ but also
${\rm Im\,}[H_{12}^2] \neq 0$, at the leptogenesis scale $M$.
Although the first condition is easily guaranteed by the running of
$M_R$ from $\Lambda_D$ to $M$, the second one requires the inclusion
of quantum corrections to the Dirac neutrino Yukawa matrix $Y$. The
renormalization group equation (RGE) for $M_R$ is, for the extended
standard (SM) and minimal supersymetric standard (MSSM)
models~\cite{Casas:1999tp},
\begin{equation}
\label{RGEMR} \frac{d M_R}{dt}=c\,(H^T M_R+M_R H)\,,\,
t=\frac{\ln\left(\mu/\Lambda_D\right)}{16\,\pi^2}\,,
\end{equation}
with $c_{{\rm SM}}=1$ and $c_{{\rm MSSM}}=2$.
%From this equation, and neglecting the running of $Y_\nu$, one has
%in the leading-log approximation $M_R(t) \propto \openone +
%(H^T+H)\, t$. It is then possible to show that the trace in
%Eq.~(\ref{J2}) is a real quantity, which leads to
%$\mathcal{J}_{2}=0$. Consequently, the $CP$-violating effects, to
%which the $CP$ asymmetries are sensitive, vanish at the decoupling
%scale.\footnote{This was first pointed out in
%Ref.~\cite{Hambye:2004jf}.} Thus, the corrections coming from the
%running of $Y_\nu$ from $\Lambda_D$ to $M$ must be taken into
%account.

In the basis where $M_R$ is diagonal, and assuming that the
charged-lepton Yukawa matrix $Y_\ell$ is diagonal, the evolution of
$Y$ and the heavy Majorana masses $M_i$ is given at one-loop by
\begin{align}
\label{RGEMRdiag} \frac{d M_i}{dt}&=2\, c\,M_i\, H_{ii}\,,
\\
\label{RGEMRdiag1} \frac{d Y}{dt}&=k\,Y+\left[ -a\,Y_\ell^{}
Y_\ell^\dagger-b\,Y Y^\dagger \right] Y + Y T \,,
\\
\label{RGEMRdiag2} \frac{d \,H}{dt}&= 2  kH- 2bH^2-2a Y^\dag
Y_\ell^{} Y_\ell^\dag Y \!+[H,T]\,,
\end{align}
where $k$ is a function of ${\rm Tr}(Y_X^{}Y_X^\dag)$ and the gauge
couplings~\cite{Chankowski:2001mx} and $[H,T]=HT-TH$. For the SM and
MSSM, the factors $a$ and $b$ are
\begin{align}
&a_{{\rm SM}}=-b_{{\rm SM}}=\frac{3}{2}\;,\;b_{{\rm
MSSM}}=3\,a_{{\rm MSSM}}=-3\,.
\end{align}
The matrix $T$ is anti-Hermitian with $T_{ii}=0$ and
\begin{equation} \label{Rmatrix}
T_{12}=\frac{2+\delta_N}{\delta_N}\, {\rm Re}\,[H_{12}]+
i\frac{\delta_N}{2+\delta_N}\, {\rm Im}\,[H_{12}]\,.
\end{equation}
Here, the parameter $\delta_N \equiv M_2/M_1-1$ quantifies the
degree of degeneracy between $M_1$ and $M_2$.

From Eq.~(\ref{Rmatrix}) on can see that if $\delta_N=0$ at a given
scale $\Lambda_D$, then the RGE in Eqs.~(\ref{RGEMRdiag1}) and
(\ref{RGEMRdiag2}) become singular, unless one imposes ${\rm
Re}\,(H_{12})=0$. This can be achieved by rotating the heavy fields
by an orthogonal transformation $O$, being the rotation angle
$\theta$ such that
\begin{equation}\label{O}
%\tan 2\theta = \frac{2\,{\rm Re}\,[H_{12}]}{H_{22}-H_{11}}\,.
\tan 2\theta = 2\,{\rm Re}\,[H_{12}]/(H_{22}-H_{11})\,.
\end{equation}
Under this transformation, $Y \rightarrow Y^\prime=YO$ and $H
\rightarrow H^\prime={Y^\prime}^\dag Y^\prime=O^\dagger H O$. It is
straightforward to show that
\begin{equation} \label{Hnurotated}
    H^\prime =\left(%
\begin{array}{cc}
  H_{11}-\Delta & i\, {\rm Im}\,[H_{12}] \\
  -i\, {\rm Im}\,[H_{12}] & H_{22}+\Delta \\
\end{array}%
\right),
\end{equation}
where $\Delta \equiv \tan \theta\, {\rm Re}\,[H_{12}]$. From
$\delta_N=M_2/M_1-1$ and Eq.~(\ref{RGEMRdiag}) one has
\begin{align}
\frac{d
\delta_N}{dt}&=2\,c\,(\delta_N+1)(H_{22}^\prime-H_{11}^\prime)\,.
\end{align}
In the limit $\delta_N \ll 1$, the leading-log approximation for
$\delta_N(t)$ can be easily found to be
\begin{equation}
\delta_N(t)\simeq 2\,c\,(H_{22}^\prime-H_{11}^\prime)\,t\,.
\end{equation}

For quasi-degenerate Majorana neutrinos the $CP$-asymmetries
generated in their decays are approximately given
by~\cite{Pilaftsis:2003gt}
\begin{equation}\label{e12d2}
\varepsilon_j=\frac{{\rm
Im}\,[\,H^{\prime\,2}_{21}]}{16\,\pi\,\delta_N\,H^{\prime}_{jj}}\!\!\left(1
+ \frac{\Gamma_i^2}{4M^2\delta_N^2}\right)^{-1}%\!\!\!\!\!\!\simeq
%\frac{{\rm
%Im}\,[\,H^{\prime\,2}_{21}]}{16\,\pi\,\delta_N\,H^{\prime}_{jj}}\,.
\!\!\!\!\,,\,j=1,2\,.
\end{equation}
Here, $\Gamma_{i}=H^{\prime}_{ii} M_i/(8\pi)$ is the tree-level
decay width of the heavy Majorana neutrino $N_i$.
\begin{figure*}\begin{center}
\begin{tabular}{ccc}
\hspace*{-0.25cm}\includegraphics[width=5.2cm]{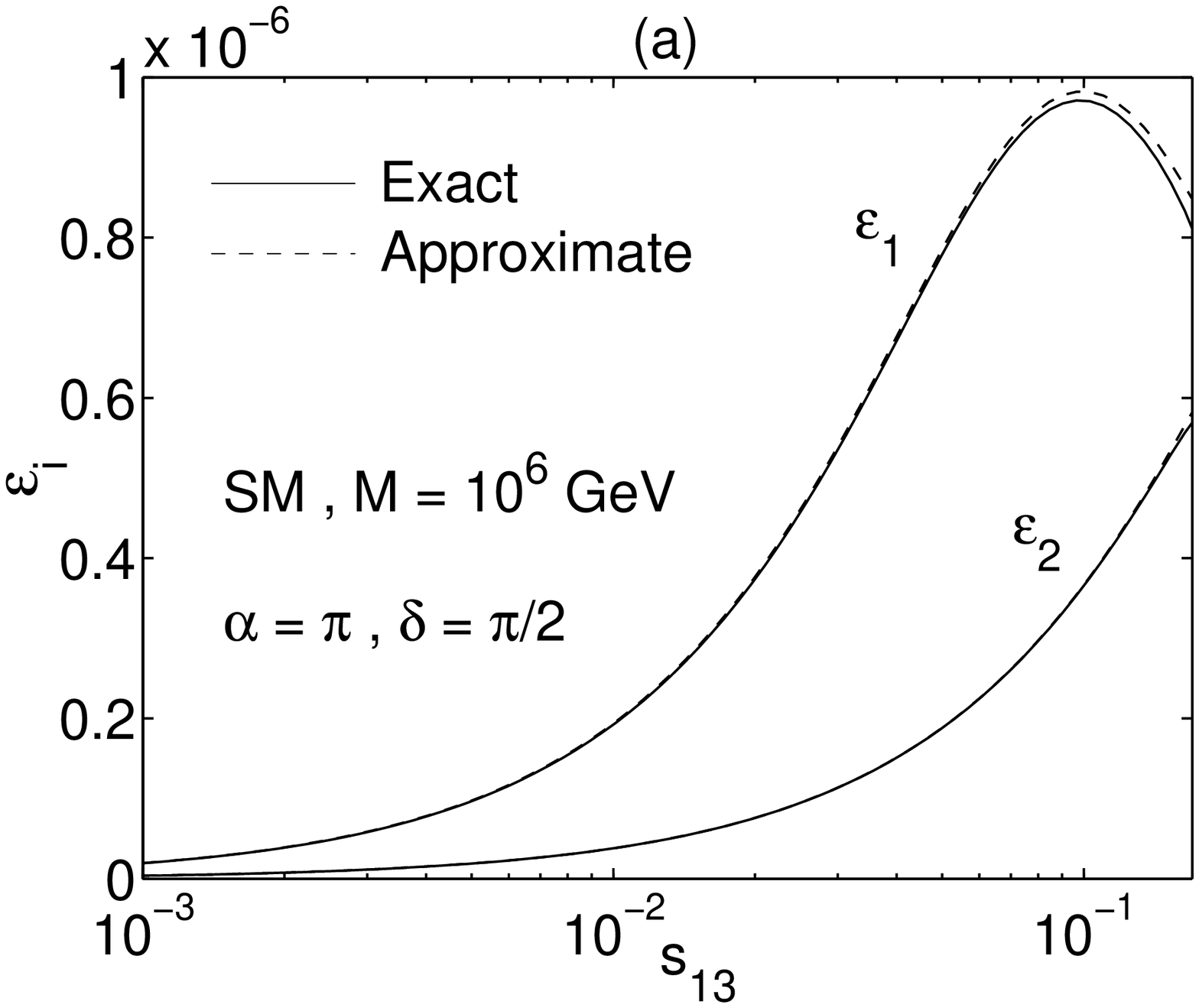}
&\hspace*{-0.25cm}\includegraphics[width=5.2cm]{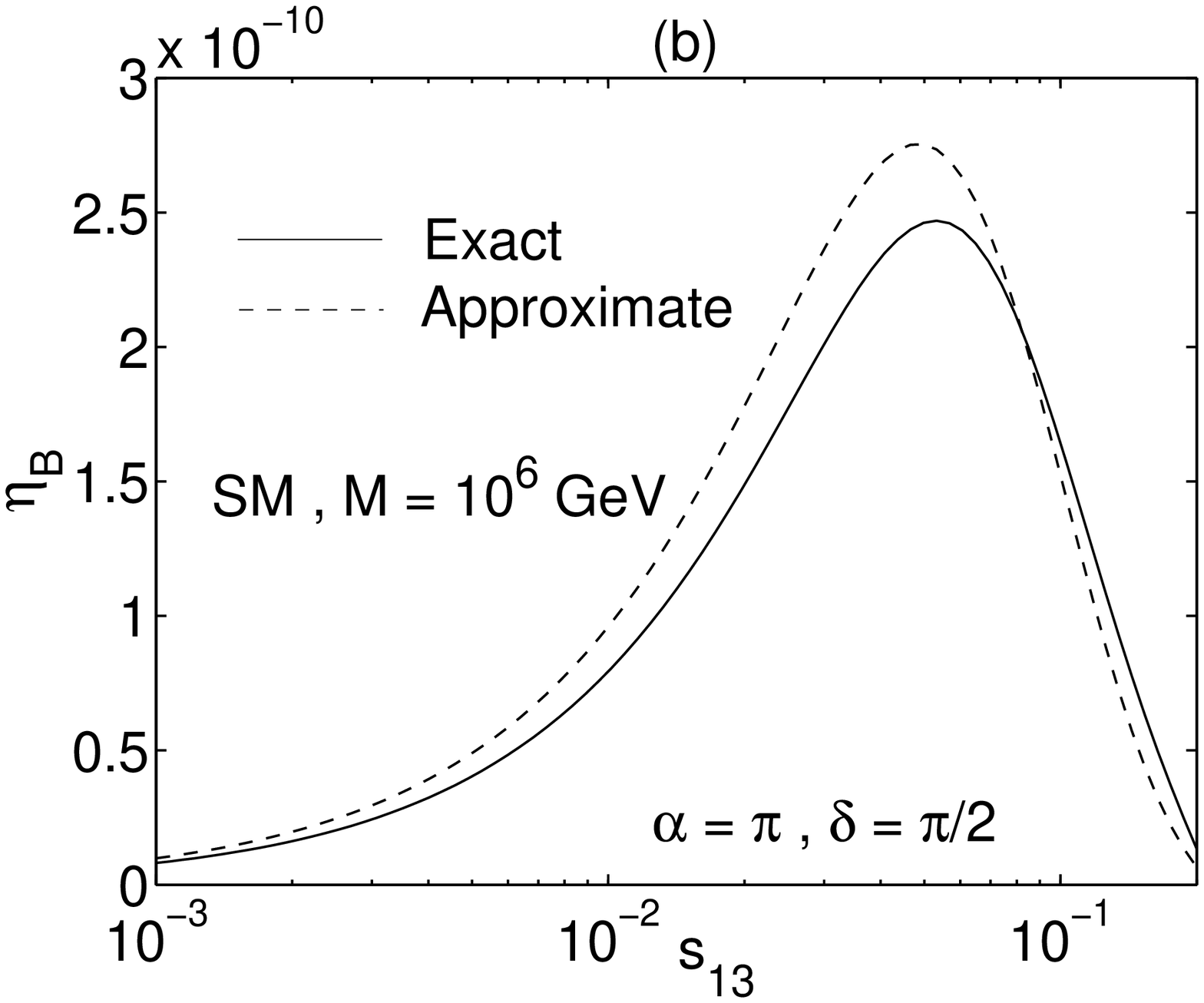}
&\hspace*{-0.25cm}\includegraphics[width=5.2cm]{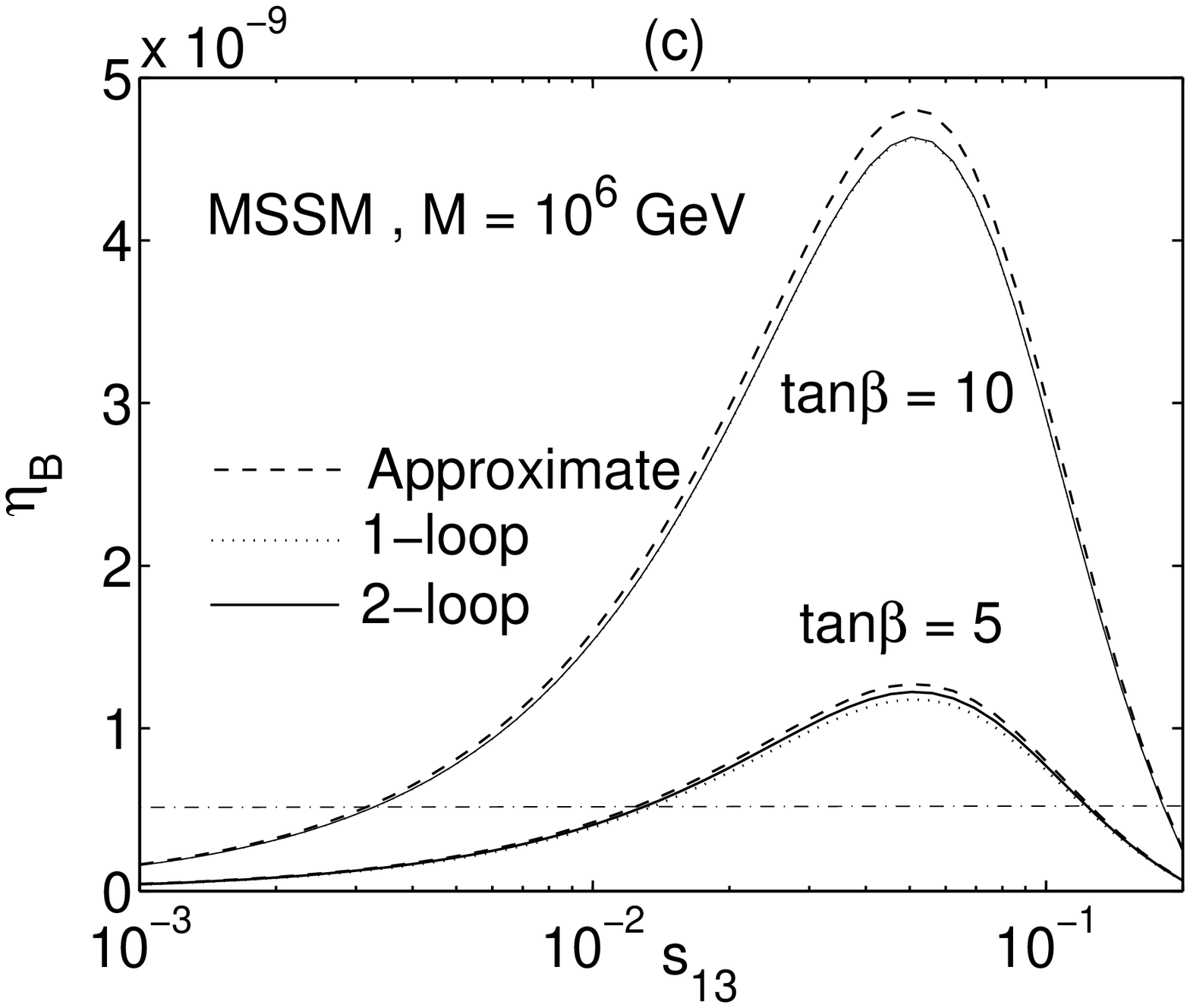}
\end{tabular}\end{center}
\vspace*{-1cm}%\caption{(a) SM $CP$-asymmetries $\varepsilon_{1,2}$
%as function of $s_{13}$ considering $\alpha=\pi$, $\delta=\pi/2$,$,
%\Lambda_D=10^{16}$~GeV and $M=10^6$ GeV. The dashed lines show the
%result of the approximations shown in Eqs.~(\ref{CPapp}) and
%(\ref{CPapp1}). (b) SM values of $\eta_B$ as function of $s_{13}$
%using the Boltzmann equations including both heavy neutrinos
%(solid-line) and using an approximation for the leptogenesis
%efficiency factor (dashed-line). (c) $\eta_B$ dependence on $s_{13}$
%in the MSSM for $\tan \beta=5,10$. The horizontal line indicates the
%mean experimental value for $\eta_B$.}
\caption{$CP$ and baryon asymmetries as functions of $s_{13}$
considering $\alpha=\pi$, $\delta=\pi/2$, $\Lambda_D=10^{16}$~GeV
and $M=10^6$ GeV (see text for more details). }
\label{fig1}\end{figure*} %
The above equation shows that
\begin{align}
\varepsilon_i(t) \propto
\text{Im}\,[H^{\prime}_{12}\,(t)\,]\,\text{Re}\,[H^{\prime}_{12}\,(t)\,]\,\,,\,\,i=1,2\,.
\end{align}
Therefore, a necessary condition to have a nonzero $CP$-asymmetry at
a given $t$ is that $\text{Re}\,[H^{\prime}_{12}\,(t)] \neq 0$.
Since $\text{Re}\,[H^{\prime}_{12}\,(0)] = 0$, one has to rely on
running effects to generate a nonzero
$\text{Re}\,[H^{\prime}_{12}]$. From Eqs.~(\ref{Rmatrix}) and
(\ref{RGEMRdiag2}) we obtain
\begin{align}\label{Reh12RGE}
\frac{d \,\text{Re}[H^\prime_{12}]}{dt} &\simeq
\left\{\frac{2\,c}{\delta_N}(H^\prime_{11}-H^\prime_{22})\,
\text{Re}[H^\prime_{12}]\right.\nonumber\\
&\left.-2\,a\,\text{Re}\,[ (Y^{\prime \dag}\,Y_\ell\,
Y_\ell^\dag\,Y^\prime)_{12}]\right\}\,.
\end{align}
Taking into account that $\text{Re}\,[H^{\prime}_{12}\,(0)] = 0$,
then
\begin{align}
\text{Re}[H^\prime_{12}(t)] \simeq
-\frac{a\,y_\tau^2}{16\pi^2}\,\text{Re}\,[Y_{31}^{\prime\ast}
\,Y_{32}^\prime]\,t\,,
\end{align}
which, in terms of the Yukawa matrix $Y$, reads
\begin{align}
\label{Re} \text{Re}[H^\prime_{12}(t)] &\simeq-
\frac{a\,y_\tau^2}{16\pi^2}\,\left\{\,\text{Re}\,[Y_{31}^\ast\,Y_{32}]
\cos 2\theta\right.\nonumber\\
&\left.+\sin2\theta\,(\,|Y_{31}|^2-|Y_{32}|^2\,)/2\,\right\}t\,.
\end{align}
The radiatively generated $\varepsilon_{1,2}$ can be computed from
Eqs.~(\ref{e12d2}) and (\ref{Re}).

In the following we will illustrate how the mechanism described
above works for a specific example. It is convenient to define the
$3\times3$ seesaw operator $\kappa$ at $\Lambda_D\,$, $\kappa=Y
\,Y^T/M$, where $Y_{ij}=y_0\,y_{ij}$ is a $3\times 2$ complex
matrix.
%\begin{equation}
%\label{seeop} \kappa=\frac{1}{M}Y \,Y^T\,\,,\,\,Y=y_0\left\lgroup
%\begin{array}{cc}
%  Y_{11} & 0 \\
%  Y_{21} & Y_{22} \\
%  Y_{31} & Y_{32}
%\end{array}
%\right\rgroup\,,
%\end{equation}
In order to reconstruct the high energy neutrino sector in terms of
the low energy parameters, we choose $y_{12}=0$. The effective
neutrino mass matrix $\mathcal{M}$ is
\begin{equation}
\label{Mnu1} \mathcal{M}=m_3\,U{\rm
diag}(0,\rho\,e^{i\alpha},1)\,U^T,\,\rho \equiv m_2/m_3\,,
\end{equation}
where $m_3$ is the mass of the heaviest neutrino and  $\alpha$ is a
Majorana phase. In the present case $m_2=\sqrt{\Delta m^2_\odot}$
and $m_3=\sqrt{\Delta m^2_a}$, where $\Delta m^2_\odot$ and $\Delta
m^2_a$ are the solar and atmospheric neutrino mass-squared
differences measured by neutrino oscillation experiments. When
necessary, we will use the best-fit values $\Delta m^2_\odot=8.1
\times 10^{-5}\,\evolt^2$ and $\Delta m^2_a=2.2\times
10^{-3}\,\evolt^2$~\cite{Maltoni:2004ei}. From these, one can
determine $\rho=\sqrt{\Delta m^2_\odot/\Delta m^2_a}$.

The matrix $U$ is the leptonic mixing matrix which can be
parametrized, as usual, in terms of three mixing angles
$\theta_{ij}$ and a Dirac-type $CP$-violating phase $\delta$. The
best fit-values for $\theta_{ij}$ are $\sin^2\theta_{12}=0.3$ and
$\sin^2\theta_{23}=0.5$, where $\theta_{12}$ and $\theta_{23}$
denote the atmospheric and solar mixing angles,
respectively~\cite{Maltoni:2004ei}. For $\theta_{13}$ we take the
$3\sigma$ bound $\sin^2\theta_{13}<0.047$~\cite{Maltoni:2004ei}.

Since the radiative corrections to $\mathcal{M}$ are negligible in
the case where its eigenvalues are hierarchical, one has
$\mathcal{M}\simeq v^2\,Y\,Y^T/M$ which implies
$y_0^2=M\,\sqrt{\dmatm}/v^2$.
%\begin{equation} \label{seesaw2}
%y_0^2=\frac{M\,\sqrt{\dmatm}}{v^2}\;,\;v \simeq 174\,\text{GeV}  \,.
%\end{equation}
%This relation leads to the following simple approximation for the
%reconstructed Dirac neutrino matrix
%\begin{align}
%&H \simeq \frac{y_0^2}{\sqrt{1+2x^2\cos \alpha+x^4}} \left\lgroup
%\begin{array}[c]{cc}
%\rho+x^2 & x\,\xi\\
%x\,\xi^\ast & 1+\rho x^2
%\end{array}
%\right\rgroup\!,\nonumber \\
%&\xi\!=\!e^{i \alpha/2}-\rho \,e^{-i\alpha/2}\,,\,x= \frac{\tan
%\theta_{13}}{\sqrt{\rho}\, s_{12}} \simeq
%\frac{s_{13}}{\sqrt{\rho}\, s_{12}}.  \label{Hnuapprox}
%\end{align}
%
In terms of the low-energy neutrino parameters, $\varepsilon_{1,2}$
take the approximate form~\cite{GonzalezFelipe:2003fi}
\begin{align}
\label{CPapp} &\varepsilon_1 \simeq -\frac{3y_\tau^2}{64\pi}
\frac{x\sqrt{\rho} \,(1+\rho)\sin
(\alpha/2)}{(1-\rho)(\rho+x^2-\Delta)}\left[c_{12}
\cos (\delta-\alpha/2)\right. \nonumber\\
& \left. + \sqrt{\rho}\, s_{12}^2\, x \cos
(\alpha/2)\right]\,,\,\varepsilon_2 \simeq
\frac{\rho+x^2-\Delta}{1+\rho x^2+\Delta}\varepsilon_1,
\end{align}
%Moreover, $\varepsilon_2$ is related with $\varepsilon_1$ through
%\begin{align}\label{CPapp1}
%\varepsilon_2 &\simeq \frac{\rho+x^2-\Delta}{1+\rho
%x^2+\Delta}\,\varepsilon_1\,,
%\end{align}
where $x=\tan \theta_{13}/(\sqrt{\rho}\, s_{12})$ and
\begin{align}
\Delta = \frac{1}{2}(1-\rho)\left[-1+x^2+\sqrt{1+2x^2\cos
\alpha+x^4} \right]\,. \nonumber
\end{align}
We use the notation $s_{ij}\equiv \sin\theta_{ij}$ and $c_{ij}\equiv
\cos\theta_{ij}$. Taking for instance $\alpha \simeq \pi$ and
$\delta \simeq \pi/2$, the $CP$ asymmetry $\varepsilon_1$ reaches
its maximum value for $x = \sqrt{\rho}\,$, as can be readily seen
from Eq.~(\ref{CPapp}). This corresponds to $s_{13} = \rho s_{12}
\simeq 0.1$ and
\begin{align}
\label{e1max}
    \varepsilon_1^{\text{max}} \simeq -\frac{3y_\tau^2\,c_{12}}{128\pi} \frac{
(1+\rho)}{(1-\rho)} \simeq -10^{-6}\,.
\end{align}

The accuracy of these approximate expressions is shown in
Fig.~\ref{fig1}.a, where the $CP$ asymmetries $\varepsilon_i$ are
plotted as functions of $s_{13}$ taking $\Lambda_D=10^{16}$~GeV,
$M=10^6$~GeV, $\delta=\pi/2$, $\alpha=\pi$ and assuming $y_\tau
=0.01$ in the analytical estimates. The solid lines correspond to
the full numerical integration of the RGE, while the dashed ones
refer to the approximations given in Eq.~(\ref{CPapp}). The
comparison of the curves shows that, for values of $s_{13} \lesssim
0.1$, $\varepsilon_2$ is suppressed with respect to $\varepsilon_1$,
in accordance with Eq.~(\ref{CPapp}). Also, the true value of
$\varepsilon_1^{\text{max}}$ agrees with Eq.~(\ref{e1max}).

The out-of-equilibrium Majorana decays are controlled by the
parameters $K_i=\Gamma_{i}/H(T=M_i)$ where $H(T)=1.66 g_\ast^{1/2}
T^2/M_P$ is the Hubble parameter, $g_\ast \simeq 107$ is the number
of relativistic degrees of freedom and $M_P = 1.2 \times
10^{19}$~GeV is the Planck mass. Considering that the entropy
remains constant while the universe cools down from $T \simeq M$ to
the recombination epoch, the baryon-to-photon ratio $\eta_B$ can be
estimated using $ \eta_B \simeq -10^{-2}\, (d_1\,\varepsilon_1 +
d_2\,\varepsilon_2)\,$, where $d_i \leq 1$ are efficiency factors
which account for the washout effects. In our case, it can be shown
that $|\varepsilon_2/K_2| \ll |\varepsilon_1/K_1|$ which leads to
$\eta_B \simeq -10^{-2}\, d_1\,\varepsilon_1$.

A simple estimate of the dilution factor $d_1$ can be obtained from
the fit $d_1 \simeq 0.6\, [\,\ln
(K_1/2)]^{-0.6}/K_1$~\cite{Kolb:vq}, where $K_1$ is, in this case,
independent of $M$ and given by
\begin{align} \label{K1app}
    K_1 \simeq \frac{44\, (\rho+x^2-\Delta)}{\sqrt{1+2x^2\cos \alpha+x^4}}\,.
\end{align}
 For $\alpha \simeq \pi,\,
\delta \simeq \pi/2$, the maximal value of the baryon asymmetry is
then attained for $x \simeq \sqrt{3\rho/(1+\rho)}/3 \simeq 0.23$ and
$s_r \simeq 0.05$. From Eq.~(\ref{CPapp}) we find $\varepsilon_1
\simeq -8\times10^{-7}$. Moreover, since Eq.~(\ref{K1app}) implies
$K_1 \simeq 11$, then $d_1 \simeq 4 \times 10^{-2}$ and
$\eta_B^{\text{max}} \simeq 3 \times 10^{-10}$, which is by a factor
of two smaller than the observed baryon asymmetry
$\eta_{B}=6.1_{-0.2}^{+0.3}\times10^{-10}$~\cite{Spergel:2003cb}. It
is worth noticing that this result is weakly dependent (apart from
renormalization effects on $y_\tau^2$) on the heavy Majorana
neutrino mass scale $M$, as can be seen from the approximate
expressions given in Eq.~(\ref{CPapp}). In Fig.~\ref{fig1}.b we
present the computation of $\eta_B$ as a function of $s_{13}$. The
dashed line refers to the result using only the decay of $N_1$ and
considering an approximation for the efficiency factor, while the
solid line has been obtained solving the full set of Boltzmann
equations, considering both the decays of $N_1$ and $N_2$.

In Fig.~\ref{fig1}.c $\eta_B$ is computed for the MSSM case.
Regarding the computation of the $CP$-asymmetries, a factor of two
has to be included in Eq.~(\ref{e12d2}) due to the presence of
supersymmetric particles in the decays. Moreover, since
$\varepsilon_{1,2} \propto y_\tau^2$, we expect an extra enhancement
factor of $(1+\tan^2\beta)$ in the MSSM respective to the SM
case~(see also Ref.~\cite{Turzynski:2004xy}). At the end, this leads
to an increase of the value of $\eta_B$. The dotted and solid lines
refer to the calculation where the $CP$-asymmetries were computed
using the RGE at one and two-loop orders, respectively, for
$\tan\beta=5,10$. The results show that the two-loop running effects
can be perfectly neglected. Moreover, the maximum of $\eta_B$ can be
far above the experimental value, which indicates that in the MSSM
there is some freedom in the choice of the $CP$-violating phases and
$s_{13}$. As in the SM case, the result based on an approximation
equivalent to the one in Eq.~(\ref{CPapp}) (dashed-lines) agree with
the exact result.

To conclude, it is worth noting that if the above mechanism is
extended to models with three heavy Majorana neutrinos, then one can
obtain values for $\eta_B$ compatible with the experiment even in
the SM case~\cite{prep}.

%\vspace*{-0.3cm}

\end{document}